\documentclass[9pt,twocolumn,twoside]{pnas-new}

\templatetype{pnasresearcharticle} 

\newcommand{\um} {~\mu\mathrm{m}}
\newcommand{\s} {~\mathrm{s}}
\newcommand{\qpar} {\mathbf{q}_\parallel}
\newcommand{\qperp} {\mathbf{q}_\perp}
\newcommand{\gammadot} {\dot{\gamma}}
\newcommand{\rd}[1]{{\color{black} #1}}

\title{Microscopic dynamics and failure precursors of a gel under mechanical load}

\author[1]{Stefano Aime}
\author[1]{Laurence Ramos}
\author[1]{Luca Cipelletti\textsuperscript{2}}

\affil[1]{L2C CNRS, Univ Montpellier, Montpellier, France}

\leadauthor{Aime}

\significancestatement{The sudden, catastrophic failure of materials under a constant mechanical load is widespread, occurring from geological scales, as in earthquakes, to biological and soft matter systems, as in protein, polymer and colloidal gels. Failure is often preceded by a long induction time, with little if any macroscopic signs of weakening, making it unpredictable. Here, we investigate the failure of a model colloidal gel under a constant load, measuring simultaneously its mechanical response and microscopic dynamics. We show that the gel undergoes a burst of microscopic plastic rearrangements that largely precede failure, undetectable by monitoring conventional 
structural quantities. The notion of dynamic precursor thus emerges as a powerful tool to understand and predict the sudden failure of solids.}

\authorcontributions{LR and LC designed the experiments. SA conducted the experiments and analyzed the data. All authors contributed to data interpretation and to writing the manuscript.}
\authordeclaration{We declare no conflict of interest.}
\equalauthors{All authors contributed equally to this work.}
\correspondingauthor{\textsuperscript{2}To whom correspondence should be addressed. E-mail: luca.cipelletti@umontpellier.fr}


\begin{abstract}
Material failure is ubiquitous, with implications from geology to everyday life and material science. It often involves sudden, unpredictable events, with little or no macroscopically detectable precursors. A deeper understanding of the microscopic mechanisms eventually leading to failure is clearly required, but experiments remain scarce. Here, we show that the microscopic dynamics of a colloidal gel, a model network-forming system, exhibit dramatic changes that precede its macroscopic failure by thousands of seconds. Using an original setup coupling light scattering and rheology, we simultaneously measure the macroscopic deformation and the microscopic dynamics of the gel, while applying a constant shear stress. We show that the network failure is preceded by qualitative and quantitative changes of the dynamics, from reversible particle displacements to a burst of irreversible plastic rearrangements.
\end{abstract}

\doi{\url{www.pnas.org/cgi/doi/10.1073/pnas.1717403115}}

\begin{document}


\maketitle
\thispagestyle{firststyle}
\ifthenelse{\boolean{shortarticle}}{\ifthenelse{\boolean{singlecolumn}}{\abscontentformatted}{\abscontent}}{}

\dropcap{M}aterial failure is ubiquitous on length scales ranging from a few nanometers, as in fracture of atomic or molecular systems~\cite{weibull1939phenomenon,celarie_glass_2003} up to geological scales, as in earthquakes~\cite{sethna_crackling_2001,ben-zion_collective_2008}. While some attempts have been made to harness failure, e.g. in order to produce new materials with a well controlled patterning~\cite{nam_patterning_2012}, material failure remains in general an unwanted, uncontrolled and unpredictable process, widely studied since the pioneering experiments on metallic wires by Leonardo da Vinci in the fifteenth century~\cite{stephen_p._timoshenko_history_1953}. Indeed, a better control of the conditions under which material failure may or may not occur and the detection of any precursors that may point to incipient failure are the Holy Grail in many disciplines, from material science~\cite{
	pradhan_crossover_2005,
	vinogradov_evolution_2012,
	amon_experimental_2013,
	koivisto_predicting_2016}
to biology~\cite{bell_models_1978,gobeaux_power_2010}, engineering and geology~\cite{
	swanson_predicting_1983, 
	mcguire_foreshock_2005, 
	wu_precursors_2006, 
	kromer_identifying_2015
	}.
Failure may occur almost instantaneously, as a consequence of an impulsive load. Often, however, it manifests itself in more elusive ways, as in the sudden, catastrophic breakage of a material submitted to a constant load, where failure may be preceded by a long induction time with little if any precursor signs of weakening. Such delayed failure has been reported in a wide spectrum of phenomena, from earthquakes~\cite{onaka_physics_2013}, snow avalanches~\cite{reiweger_load-controlled_2010} and failure in biomaterials~\cite{bell_models_1978,gobeaux_power_2010} to the sudden yielding of crystalline~\cite{weibull1939phenomenon
} solids, composite materials~\cite{nechad_creep_2005,koivisto_predicting_2016} and amorphous systems~\cite{preston_mechanical_1942}, including viscoelastic soft materials~\cite{siebenburger_creep_2012,sentjabrskaja_creep_2015-1}, such as adhesives~\cite{sancaktar_material_1985} and network-forming materials~\cite{bonn_delayed_1998,
	poon_delayed_1999,
	skrzeszewska_fracture_2010,
	sprakel_stress_2011,
	leocmach_creep_2014}.

Delayed failure typically involves creep during the induction time, the sub-linear \rd{(e.g. power-law)} increase of sample deformation under a constant load. The microscopic origin of creep is well understood for crystalline solids, where it is attributed to defect motion~\cite{e.n._andrade_viscous_1910,miguel_dislocation_2002}. Power-law creep is also widespread in amorphous materials, but its microscopic origin remains controversial: it has been attributed to the accumulation of irreversible, plastic rearrangements~\cite{siebenburger_creep_2012,
	sentjabrskaja_creep_2015-1,
	nechad_creep_2005,
	coussot_aging_2006,
	caton_plastic_2008}, to linear viscoelasticity~\cite{balland_power_2006,gobeaux_power_2010,leocmach_creep_2014},
or to a combination of both~\cite{
	jagla_creep_2011}, with different authors holding contrasting views on similar systems ~\cite{nechad_creep_2005,jagla_creep_2011}. Crucially, a detailed understanding of the creep regime holds the promise of unveiling the origin of the sudden failure of the material, potentially revealing any precursor signs of failure, which are difficult to detect by monitoring macroscopic quantities, such as the deformation rate~\cite{koivisto_predicting_2016}, or mesoscopic, coarse-grained shear velocity maps~\cite{leocmach_creep_2014}. Clearly, investigations of the evolution of the microscopic structure and dynamics under creep are required, which are however very scarce to date and essentially restricted to numerical works~\cite{landrum_delayed_2016}.

Here, we address these questions by studying the microscopic dynamics of a soft solid submitted to a constant shear stress, using a unique custom-made apparatus~\cite{aime_stress-controlled_2016,tamborini_multiangle_2012} that couples stress-controlled rheology to small-angle static and dynamic light scattering \rd{(see the SI for details on the sample and setup geometry)}. We focus on a gel made of attractive colloidal particles, a model system for network-forming soft solids, which are ubiquitous in soft matter~\cite{zaccarelli_colloidal_2007} and in biological materials~\cite{storm_nonlinear_2005}. Initially, particles in the gel network undergo both affine (as in ideal elastic solids) and non-affine displacements, but all displacements are fully reversible. Thus, the initial regime of creep is not due to plasticity, but rather to the complex viscoelastic response of the gel network. At larger strains, by contrast, the dynamics are due to irreversible plastic rearrangements that progressively weaken the network, eventually leading to the gel failure. Strikingly, this plastic activity does not increase steadily until failure, but rather has a non-monotonic behavior, peaking thousands of seconds before the macroscopic rupture. Our work thus establishes the notion of dynamic precursor as a powerful tool to understand and predict sudden material failure.

\begin{figure}
\centering
\includegraphics[width=\linewidth]{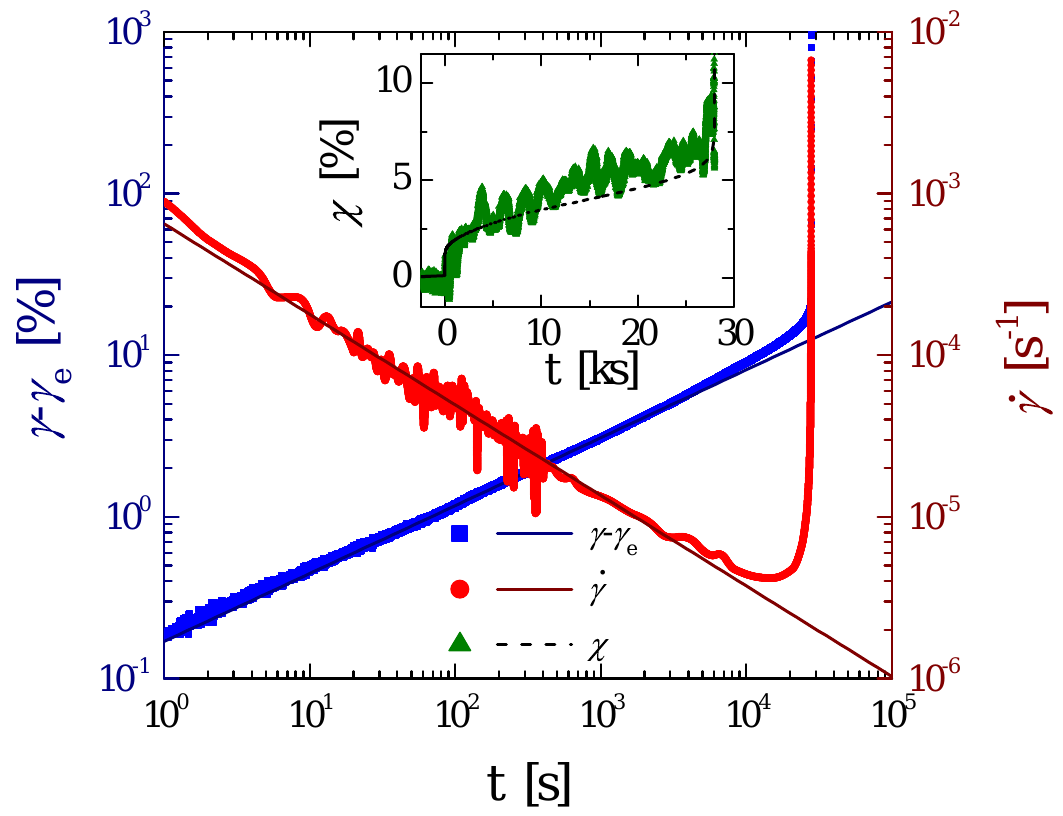}
\caption{\textbf{Mechanical response and structure evolution of a colloidal gel during creep.} Main plot: deformation in excess of the elastic jump $\gamma_e = 4.8\%$ (blue squares, left axis) and shear rate (red circles, right axis) following a step shear stress of amplitude $\sigma_0=240$ Pa, applied at time $t=0$. Lines: power law fits to the data in the initial creep regime ($1~\mathrm{s} \le t \le 10^4~\mathrm{s}$), yielding an exponent $\alpha=0.43 \pm 0.01$ in the generalized Maxwell viscoelastic model. Inset: anisotropy $\chi$ of the scattered intensity as a function of $t$, for $q_\perp=q_\parallel=2.6\um^{-1}$. Triangles: data for the creep test. Line: anisotropy as obtained from $\chi = k \gamma\rd{(t)}$, with the proportionality coefficient $k = 0.26$ determined in independent oscillatory experiments in the linear regime. Solid and dashed lines correspond to the linear regime and to an extrapolation in the non-linear regime, respectively.}
\label{fig:shrate_fit}
\end{figure}

The gel is formed \textit{in situ} by triggering the aggregation of an initially stable suspension of silica nanoparticles via an enzymatic reaction (see Methods). The nanoparticles have radius $a = 26~\mathrm{nm}$ and occupy a volume fraction $\varphi = 5\%$. Gelation occurs within 3h, resulting in a network formed by fractal clusters with typical size $\xi \sim a \varphi^{1/(d_f-3)} \sim 0.5\um$ ~\cite{carpineti_transition_1993,manley_gravitational_2005} and fractal dimension $d_f = 2$. All experiments are performed at least 48h after gelation, when the gel viscoelastic properties don't evolve significantly with sample age. Under a constant load, the gel exhibits delayed failure, a feature reported for many network-forming systems~\cite{bonn_delayed_1998,
	poon_delayed_1999,
	skrzeszewska_fracture_2010,
	sprakel_stress_2011,
	leocmach_creep_2014}.
Figure 1a demonstrates delayed failure for our gel, by showing the time evolution of the shear strain $\gamma$ and of the strain rate $\gammadot$ upon imposing a constant stress $\sigma_0=240$ Pa at time $t=0$. On time scales shorter than those shown in Fig. 1a ($t< 1$ s), the gel responds elastically: $\gamma$ jumps to an elastic shear deformation \rd{$\gamma_e \sim 4.8\%$}, corresponding to a shear modulus \rd{$G = \sigma_0/\gamma_e = 5000$ Pa}, consistent with the low-frequency elastic modulus $G'$ measured in oscillatory rheology tests (See Methods). Following the elastic jump, $\gamma$ grows sublinearly: both the deformation in excess of the elastic response, $\gamma - \gamma_e$, and the shear rate follow power laws, well accounted for by a generalized, or fractional, Maxwell viscoelastic model~\cite{jaishankar_power-law_2013}, $\gamma(t)-\gamma_e = \gamma_e \Gamma^{-1}(\alpha) (t/\tau_{FM})^\alpha$, with $\alpha=0.43 \pm 0.01$, $\tau_{FM} \gtrsim 10^5~\mathrm{s}$ and $\Gamma(x)$ the Gamma function. Remarkably, this creep regime extends over more than four decades in time, until the gel abruptly fails at $t \approx 2.8 \times 10^4$ s, as signalled by the sharp upturn of both $\gamma$ and $\gammadot$.

To investigate the relationship between the sudden macroscopic failure of the gel and its microscopic evolution, we inspect static and dynamic light scattering data collected simultaneously to the rheology measurements (see Fig. 1b~\cite{aime_stress-controlled_2016}). Light scattering probes density fluctuations as a function of wavevector $\mathbf{q}$ : $I(\mathbf{q}) \propto \sum_{j,k} \exp[-i\mathbf{q}\cdot(\mathbf{r}_j-\mathbf{r}_k)]$, with $I$ the scattered intensity, $\mathbf{r}_j$ the position of the $j$-th particle, and $\mathbf{q}$ the scattering vector (see Methods). We use a custom-designed small angle setup~\cite{tamborini_multiangle_2012} based on a CMOS detector, allowing measurements on length scales $\sim \pi/q$ in the range $0.8\um -10\um$, comparable to or larger than the cluster size $\xi$. At rest, the scattering pattern depends only on the magnitude of $\mathbf{q}$, since the gel is isotropic. During creep, the $q$ dependence of the scattered intensity hardly changes, indicating that the gel structure is fundamentally preserved until sample failure. However, a small anisotropy develops in the static structure factor, similar to that observed for other sheared soft solids~\cite{mohraz_orientation_2005-1}. 
We quantify this asymmetry by $\chi(q)=\left[I(\mathbf{q}_\parallel) - I(\mathbf{q}_\perp)\right]/\left[I(\mathbf{q}_\parallel) + I(\mathbf{q}_\perp)\right]$, with $|\qpar| = |\qperp|$ and where $\parallel$ and $\perp$ refer to orientations of the scattering vector parallel and perpendicular to the shear direction, respectively. The inset of Fig. 1a shows \rd{the time dependence of $\chi$. We find that the asymmetry follows the same trend as $\gamma(t)$, i.e. that $\chi$ is proportional to $\gamma$ throughout the  whole experiment, up to failure. Moreover, the proportionality coefficient is the same as that measured in independent oscillatory experiments in the linear, reversible regime.} Thus, structural quantities \rd{simply} reflect the macroscopic shear deformation, without providing additional information on the fate of the gel.

\begin{figure*}[tbhp]
\centering
\includegraphics[width=.8	\textwidth, clip]{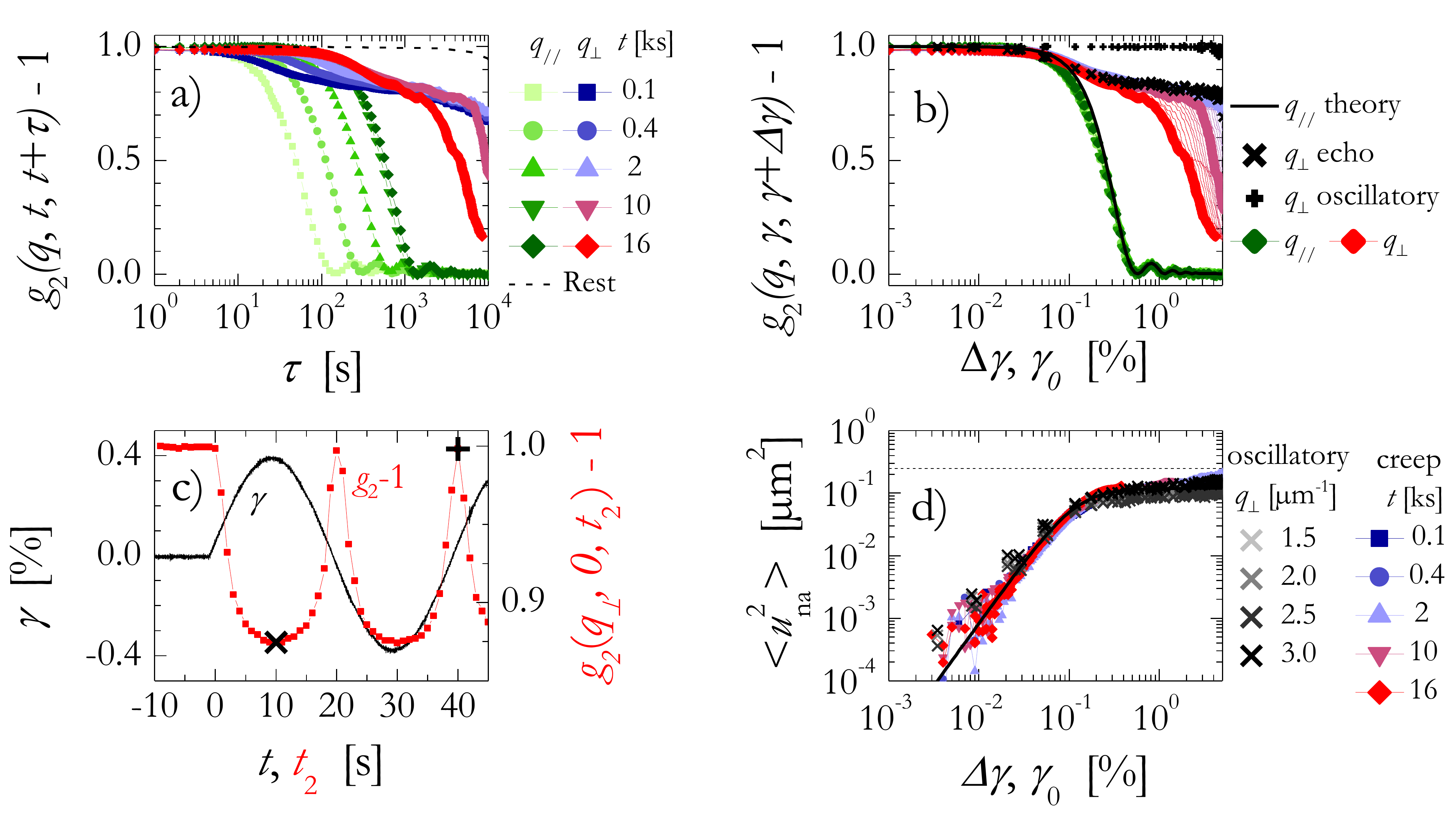}
\caption{\textbf{Microscopic dynamics of the gel during creep.} a): Time correlation functions, $g_2-1$, measured in the $q_\parallel$ (green) and $q_\perp$ (blue to red shades) directions, as a function of time lag $\tau$, for $q=3.1\um^{-1}$ and for representative times $t$ after applying a stress step. Black dashed line: spontaneous isotropic dynamics measured on the same sample, but at rest. b): Solid symbols: same data as in a), plotted as a function of the strain increment $\Delta \gamma$. Additional data sets for intermediate $t$ are shown as small symbols. Line: $g_2-1$ calculated assuming a purely affine deformation. $+$, $\times$: data collected in independent oscillatory experiments following the protocol shown in c). $\times$: maximum decorrelation during a shear cycle of amplitude $\gamma_0$; $+$: correlation echo after a full cycle. c): Shear deformation and correlation function during a shear cycle of amplitude \rd{$\gamma_0 = 0.4\%$}. $\times$ and $+$ indicate the correlation values reported in b).
d): Non-affine mean square displacement \textit{vs} strain increment during creep (solid symbols) and \textit{vs} $\gamma_0$ in oscillatory experiments. The dashed line corresponds to the \rd{squared} cluster size. The solid line is a fit to the data using $<u^2_{na}> = u^2_{\infty}\frac{\gamma_0^2}{\gamma_0^2 + \gamma_c^2}$, with \rd{$u^2_{\infty} = 0.12 \um^2$} and \rd{$\gamma_c = 0.12\%$}.}
\label{fig:creep_g2m1}
\end{figure*}

We now show that the microscopic dynamics are a much more sensitive probe of the gel evolution, unveiling dramatic plastic events that weaken the network thousands of seconds before its macroscopic failure. We measure the two-time intensity correlation function  $g_2(\mathbf{q}, t_1, t_2) - 1 \propto |g_1(\mathbf{q}, t_1, t_2)|^2$, with $g_1$ the field correlation, or intermediate scattering, function~\cite{berne_dynamic_1976} (see Methods).
The correlation function is measured simultaneously for several $\mathbf{q}$ vectors; Fig. 2a shows representative $g_2-1$ measured at various times $t_1$ during the creep, for $q = 3.1\um^{-1}$. As for the static intensity, we analyze data separately for scattering vectors parallel and perpendicular to the shear direction. The curves for $\mathbf{q}_\parallel$ (green symbols in Fig. 2a) exhibit a full decay on timescales that grow during creep, eventually reaching $\approx 10^3\s$. The behavior for $\mathbf{q}_\perp$ is more complex: initially, $g_2-1$ decays to a quasi-plateau, while a two-step relaxation leading to an almost complete decorrelation is seen at later times. Throughout the experiment, the dynamics along $\mathbf{q}_\parallel$ are faster than for $\mathbf{q}_\perp$. This can be understood by decomposing the particle displacement 
in its affine and non-affine components: $\mathbf{r}(t_2) - \mathbf{r}(t_1) =  \mathbf{u}_{0}(t_1,t_2) + \mathbf{u}_{na}(t_1,t_2)$. For a particle with coordinate $z$ in the direction of the shear gradient, the affine component is $\mathbf{u}_{0} = (\gamma_2-\gamma_1)z \hat{\mathbf{e}}_\parallel$, with $\gamma_{i} = \gamma(t_{i})$ and $\hat{\mathbf{e}}_\parallel$ the unit vector parallel to the shear direction. Because $\hat{\mathbf{e}}_\parallel \cdot \qperp = 0$, correlation functions measured for
$\mathbf{q}_\perp$ are only sensitive to $\mathbf{u}_{na}$, the non-affine component of the displacement, while the decay of $g_2(\qpar, t_1,t_2)-1$ reflects both affine and non-affine motions, resulting in a faster relaxation. In principle, additional contributions to $g_2-1$ may also stem from the spontaneous, thermally activated dynamics of the gel~\cite{cipelletti_universal_2000}. However, here this contribution is negligible (see dashed line in Fig. 2a). Thus, the microscopic dynamics is only related to the shear deformation: this suggests to analyze the dynamics as a function of strain increment, rather than time delay.

Figure 2b shows the same data as in Fig. 2a, replotted versus the strain increment $\Delta \gamma =\gamma_2 - \gamma_1$. A remarkable collapse is seen for the $\qpar$ data, independent of $\gamma_1$. This indicates that motion in the $\hat{\mathbf{e}}_\parallel$ direction is dominated by affine displacements, for which $\mathbf{u}_{0}$ is proportional to $\Delta \gamma$, regardless of the cumulated strain. We confirm this interpretation by calculating $g_2(\qpar)-1$ for a purely affine shear deformation (see Methods): the result (line in Fig. 2b) is indeed very close to the data, ruling out sample slip or shear banding, which would result in significant deviations of $g_2-1$ with respect to its theoretical form. Note however that the data lay slightly below the theoretical curve, showing that a small non-affine component must also be present. Non-affine motion is better resolved by inspecting the $\mathbf{q}_\perp$ correlation functions, which are insensitive to the affine component.

\rd{In the following, we thus focus on the microscopic dynamics in the direction perpendicular to shear, which probes only non-affine displacements.  We shall first discuss the linear viscoelastic regime, where we will show that microscopic displacements are fully reversible, and then the non-linear regime, where irreversible, plastic rearrangements come into play, ultimately causing the gel failure. In the initial regime,} $\gamma-\gamma_e \le 4 \%$ ($t \le 2000\s$, blue shades in Fig. 2b), all data collapse onto a master curve, exhibiting a decay to a quasi-plateau. This collapse is remarkable and sheds light on the nature of the non-affine deformation observed in the initial regime of the creep. For an ideal solid, the displacement under a shear deformation is purely affine. Non-affine displacements indicate a departure from this ideal behavior, which may stem from two different physical mechanisms: elastic, reversible response, but with spatial fluctuations of the elastic modulus~\cite{basu_nonaffine_2011,
	didonna_nonaffine_2005,
	leonforte_inhomogeneous_2006,zaccone}, or plastic, irreversible rearrangements~\cite{falk_dynamics_1998}. The fact that $g_2(\qperp,\gamma_1, \gamma_2)-1$ is independent of the cumulated strain suggests that no plastic events occur in the initial creep regime. We test this hypothesis by measuring reversibility at the microscopic level in separate oscillatory shear experiments, following the ``echo'' protocol of Refs.~\cite{hebraud_yielding_1997,
	petekidis_rearrangements_2002,
	leheny_rheo-xpcs_2015}.
As shown in Fig. 2c, the sample is submitted to a sinusoidal deformation, $\gamma(t) = \gamma_0 \sin(\omega t)$. The overlap between the initial microscopic configuration and a sheared one is quantified by $g_2(\qperp,t=0, t_2)-1$. We focus on the maximum of $g_2-1$ at the correlation echo, after one full cycle (`$+$' in Fig. 2c), and on its first minimum, ($t_2 = 10$ s, `$\times$' in Fig. 2c). By repeating the measurements for several $\gamma_0$, we obtain the plus and cross symbols displayed in Fig. 2b, for the maximum and the minimum level of correlation, respectively. Up to $\gamma = 4\%$, we find no loss of correlation upon \rd{setting back to zero the macroscopic deformation}; furthermore, the crosses follow the same master curve as the creep data. This confirms that for $t \le 2000\s$ the creep is not due to plastic rearrangements, but rather to the slow, fully recoverable deformation of the elastically heterogeneous network, at fixed connectivity. To characterize the strain dependence of non-affinity, we extract the non-affine mean squared displacement $<u^2_{na}>$ via $g_2(\qperp,\Delta \gamma)-1 = \exp(-\qperp^2\left <u^2_{na}(\Delta \gamma)\right >/3)$, the analogous in the strain domain of the usual relationship between correlation functions and mean squared displacement in the low $q$ limit~\cite{berne_dynamic_1976}. As seen in Fig. 2d, data collected at various $q$ collapse on the same curve, confirming the $q^2$ scaling. Initially, $<u^2_{na}>$ grows as $\Delta \gamma^2$, eventually saturating to a plateau $\approx 0.1\um^2$. The quadratic dependence of $<u^2_{na}>$ on strain is the analogous of ballistic dynamics in the time domain; it is the signature of elastic response in an heterogeneous medium~\cite{didonna_nonaffine_2005}, and was recently reported for a polymer network~\cite{basu_nonaffine_2011}. The non-affine displacement saturates at a value close to the cluster size (dotted line in Fig. 2d), consistent with the \rd{physical picture} that the gel structure cannot be remodelled on length scales larger than the network mesh size without changing its connectivity, i.e. without any plastic rearrangements.

We now turn to the $\gamma - \gamma_e > 4 \%$ regime ($t > 2000\s$, red shades in Fig. 2b), where $g_2-1$ exhibits a two-step relaxation. The initial decay overlaps with that observed at early times, due to reversible non-affine deformation. The final decay of $g_2(\qperp)-1$ indicates additional dynamics, which lead to the relaxation of density fluctuations on length scales comparable to or larger than $\xi$. In this regime, the macroscopic deformation is not recovered upon releasing the applied stress: we thus attribute the additional dynamics to irreversible plastic rearrangements. To investigate the evolution of plastic dynamics during creep, we show in Fig. 3 the intensity correlation function for several scattering vectors and a fixed strain increment $\Delta \gamma = 5\%$, versus cumulated strain. This quantity is directly related to the amount of plastic rearrangements occurring over $\Delta \gamma$. Strikingly, all curves exhibit a negative peak, indicating that the gel undergoes a burst of plastic activity for $13\% \lesssim \gamma \lesssim 22\%$. Remarkably, the minimum of $g_2-1$ occurs at $\gamma \approx 17\%$ ($t = 1.9 \times 10^4\s$), as much as 9000 s before the gel fails, for \rd{$\gamma \approx 30\%$}. Figure 3 reveals that the minimum is more pronounced for the largest $\qperp$ vectors. Thus, the burst of plastic activity is better seen when probing the dynamics on small length scales, which suggests that macroscopic quantities should be less sensitive to such burst. This is indeed the case for the macroscopic strain: as seen in Fig. 1a, only a slight deviation from the sublinear creep is seen around $t = 1.9 \times 10^4\s$, at the burst maximum. The strain rate is a more sensitive quantity: Fig. 3 shows that the onset of plasticity coincides with the departure of $\gammadot$ from its power law behavior in the linear regime (dotted line), thus establishing a direct connection between microscopic dynamics and macroscopic creep \rd{(see the SI for a detailed comparison between mechanical and dynamical signatures of plasticity)}.

To quantify the microscopic plastic activity during creep, we develop a simple model for dynamic light scattering under time-varying conditions. Using strain as the relevant variable and focussing on the dynamics along $\qperp$, we assume that reversible and plastic displacements are uncorrelated processes, leading to the factorization $g_1(\qperp,\gamma_1,\gamma_2) = R(\qperp,\Delta \gamma)P(\qperp,\gamma_1,\gamma_2)$,  with $R$ and $P$ the contributions due to reversible and plastic displacements, respectively, and where $R$ only depends on the strain increment, as indicated by the experiments. For a stationary process, a general form that captures well different kinds of dynamics is $g_1 = \exp[-f(q)(A\Delta \gamma)^p]$. The initial decay of the correlation function in the linear, reversible regime discussed in reference to Fig. 2 is an example of this functional form, with $f\sim q^2$, $p = 2$, and $A$ the constant, non-affine root-mean square particle displacement per unit strain increment. We generalize this form by expressing $P(\qperp,\gamma_1,\gamma_2)$ as a function of a strain-dependent plastic activity per unit strain increment, $A(\gamma)$:
\begin{equation}
g_1(\qperp,\gamma_1,\gamma_2)= R(\qperp,\Delta \gamma)\exp\left[-f(\qperp)\left(\int_{\gamma_1}^{\gamma_2} A(\gamma)\mathrm{d}\gamma\right)^p\right] \,.
\label{eq:P}
\end{equation}

\begin{figure}
\centering
\includegraphics[width=\linewidth]{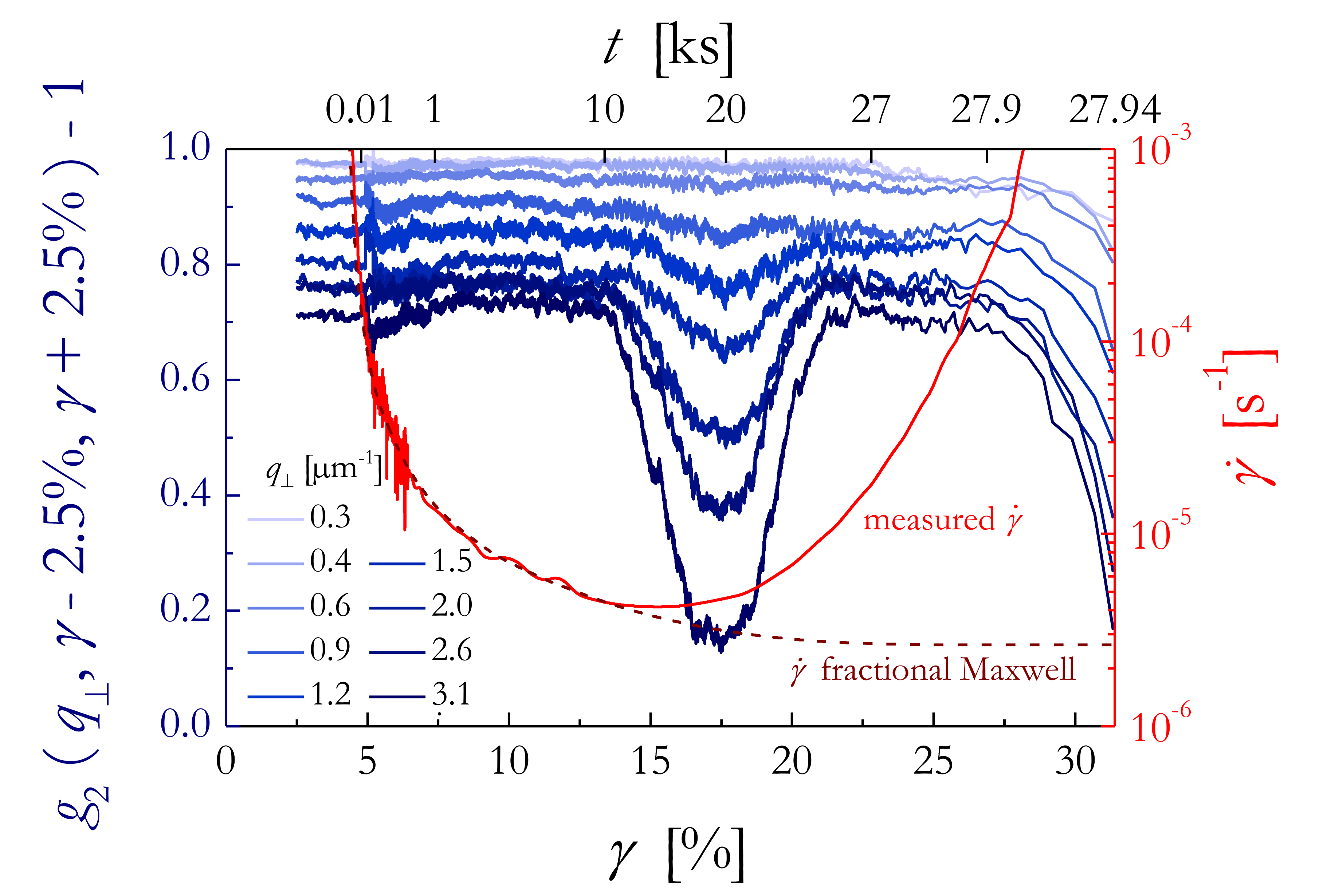}
\caption{\textbf{Microscopic dynamics signals the onset of plasticity.} $g_2(\qperp, \gamma-\Delta \gamma/2,\gamma+\Delta \gamma/2)-1$ vs $\dot{\gamma}$  Right axis, red line: macroscopic deformation rate $\dot{\gamma}$ \textit{vs} cumulated shear deformation during creep. Dashed line: generalized\rd{, or fractional,} Maxwell model. Left axis, blue curves: microscopic non-affine dynamics over a fixed strain increment $\Delta \gamma = 5\%$, for various $\qperp$ as indicated by the labels.}
\label{fig:fapres}
\end{figure}

We extract $f(\qperp)^{1/p}A(\gamma)$ from the experimental data using Eq.~(\ref{eq:P}) and assuming that $R(\qperp,\Delta \gamma)$ is the same as in the reversible regime of creep (see Methods for details). The results are shown in Fig. 4a, for several $\qperp$ vectors. Consistent with the findings for the correlation function at a specific strain increment, Fig. 3, the plastic activity per unit strain exhibits a non-monotonic behavior, with a peak centered around $\gamma = 17.1 \%$. The height of the peak strongly increases with $q$. According to our model, this is due to the $q$ dependence of the prefactor $f$, since the plastic activity $A$ is a quantity intrinsic to the gel and is thus independent of the probed length-scale. We test this assumption by plotting in Fig. 4b the plastic activity scaled by its peak value, $A_0$. An excellent collapse is seen for data spanning a factor of 5 in $q$ vectors, thereby confirming the soundness of the model.

\begin{figure}
\centering
\includegraphics[width=.9\linewidth]{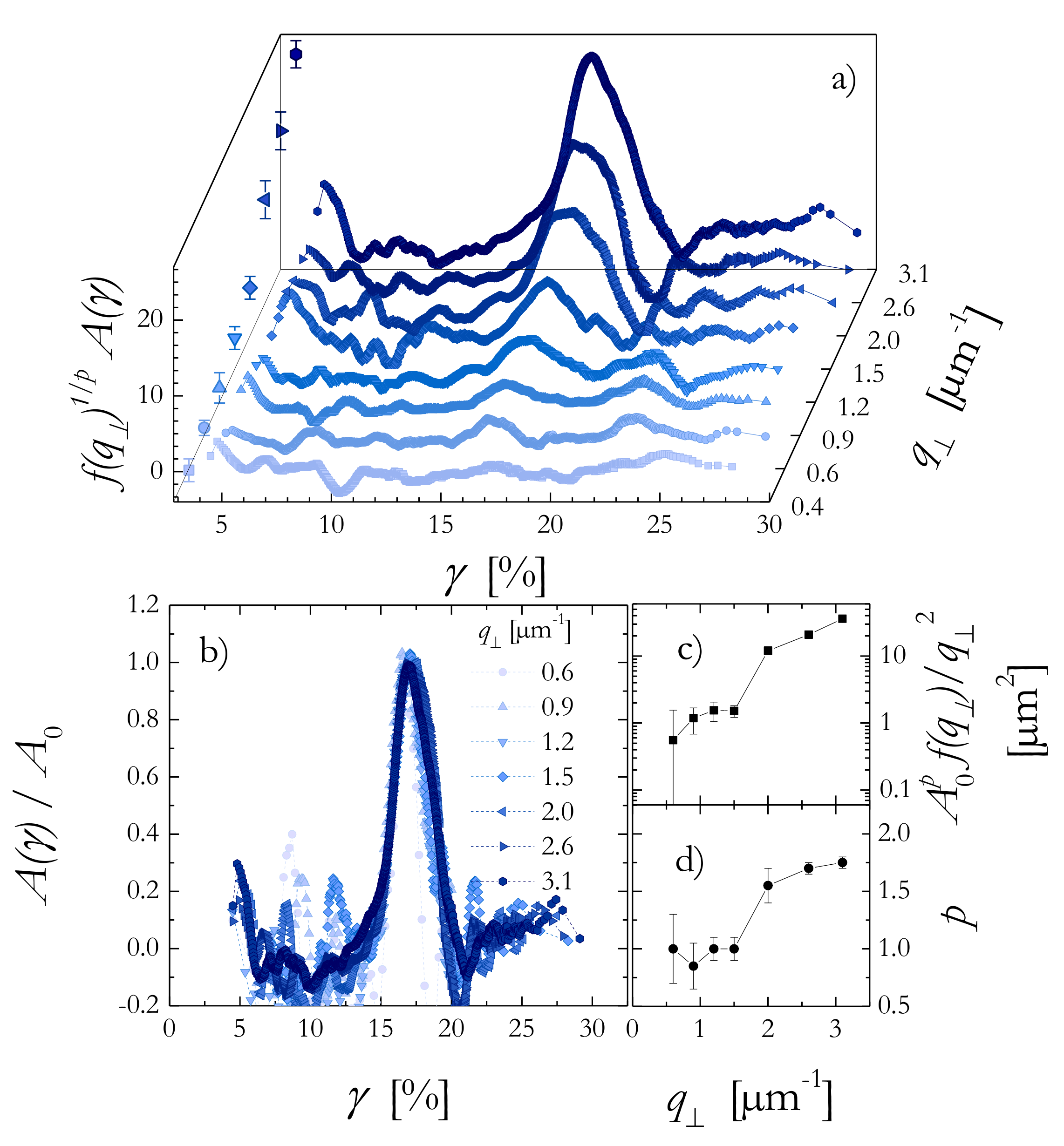}
\caption{\textbf{Plastic activity as revealed by the microscopic dynamics.} a): Plastic activity per unit strain increment \textit{vs} cumulated strain, for various $\qperp$. The error bars on the left indicate the uncertainty resulting from averaging data collected at different $\Delta \gamma$. b): Collapse of the plastic activity measured at various $\qperp$, as indicated by the labels. c): $q$ dependence of the plastic dynamics. Data are normalized by $q^2$, the behavior expected for diffusive dynamics. d): Exponent $p$ characterizing the plastic dynamics (see Eq.~\ref{eq:P}), where $p=1$ corresponds to diffusive dynamics.}
\label{fig:dofq}
\end{figure}

To gain insight into the nature of the plastic dynamics, we inspect the $q$ dependence of the prefactor $f(\qperp)A_0^p$ and of the exponent $p$. For $q_{\perp} \le 1.5\um$, the prefactor scales as $q^2$ (see Fig. 4c, where the $q^2$ dependence has been factored out) and $p \approx 1$. Under these conditions, the contribution of plasticity to the decay of the correlation function over a small strain increment reads $P(\qperp,\gamma,\gamma+\Delta \gamma) = \exp[-q_{\perp}^2 D(\gamma)\Delta \gamma]$, with $D(\gamma) \sim A(\gamma)$ a strain-dependent, but $q$-independent, diffusion coefficient. Thus, in the low $q$ regime the plastic dynamics are diffusive, since $P$ is the analogous in the strain domain of the usual diffusive dynamics in the time domain, for which $g_1 = \exp[-q^2D\tau]$~\cite{berne_dynamic_1976}. A change of the plastic dynamics occurs beyond $q_{\perp}^* \approx 2 \um^{-1}$, corresponding to a length scale $\pi/q_{\perp}^* \approx 1.6\um$, slightly larger than the cluster size. For $q_{\perp} \ge q_{\perp}^*$, $fA_0^p$ grows sharply, increasingly departing from the $q^2$ scaling. Concomitantly, the $p$ exponent grows up to $p \approx 1.75$ at the largest probed $q$ vectors (Fig. 4c), approaching $p=2$, the exponent characterizing ballistic dynamics. The emerging picture is that of plasticity consisting of irreversible rearrangements, most likely due to bond rupture. On small length scales (large $q$), the dynamics are strongly $q$ dependent and are dominated by local motion associated with such rearrangements. On larger length scales (smaller $q$), the contribution of many events add up, leading to a diffusive decay of density fluctuations. These events progressively weaken the network,  eventually leading to its catastrophic failure.

The experiments reported here unveil the complex evolution of the microscopic dynamics during the creep of a colloidal gel, from reversible non-affine motion due to the heterogeneous gel structure, to a burst of plastic rearrangements that irreversibly weaken the network, \rd{providing a microscopic signature of the onset of plasticity. Remarkably, this dynamic precursor occurs midway through the creep. While further theoretical work will be needed to fully understand the origin of the precursor and its temporal location,
we emphasize that its occurrence allows one to predict the ultimate fate of the network thousands of seconds before its catastrophic rupture.} Ongoing experiments in our group reveal that similar dramatic changes of the microscopic dynamics \rd{largely} precede failure in a variety of mechanically driven materials, from polymer gels to elastomers and semicrystalline polymers. The notion of \textit{dynamic precursor} therefore emerges as a powerful concept to understand and predict material failure.

\matmethods{
\subsection{Enzyme-induced aggregation and gel formation}
The gel results from the aggregation of a suspension of silica particles (Ludox TM50, from Sigma Aldrich, diameter $a=26$ nm as determined by small angle neutron scattering, SANS), dispersed at a volume fraction $\varphi=5$ \%  in an aqueous solvent containing urea at $1$ M. Particle aggregation is triggered by increasing \textit{in situ} the ionic strength of the solvent, thanks to the hydrolysis of urea into carbon dioxide and ammonia, a reaction catalyzed by an enzyme (Urease U1500-20KU, from Sigma Aldrich, $35$ U/ml)~\cite{wyss_small-angle_2004}, whose activity depends on temperature $T$. The suspension is prepared at $T \approx 4^\circ \mathrm{C}$ and brought at room temperature after loading the cell, thereby activating the enzyme and initiating aggregation. The sol-gel transition occurs $ \approx 3$ hours after loading the sample in the shear cell. The fractal dimension of the gel network, $d_f = 2$, has been determined from independent SANS measurements on a gel prepared following the same protocol.

\subsection{Oscillatory rheology in the linear regime}
We characterize the mechanical properties of the gel by measuring the frequency-dependent elastic and loss moduli ($G'(\omega)$ and $G''(\omega)$, respectively) in the linear regime ($\gamma_0 = 0.1 \%$) using a commercial rheometer (MCR502 by Anton Paar). Over the range $10^{-3} ~\mathrm{rad~s}^{-1} \le \omega \le 10 ~\mathrm{rad~s}^{-1}$, $G'$ is essentially flat and $G'' \sim \omega^{-0.4}$, a behavior consistent with the fractional Maxwell model that accounts for the gel creep. The gel slowly ages: $G'$ increases as $t^{1/3}$ and the \rd{characteristic} relaxation time $\tau_{FM}$ obtained from the fractional Maxwell model increases linearly with $t$. We let the gel age for 48h before running a creep experiment, such that during the \rd{duration} of one experiment (typically a few hours) the viscoelastic properties of the gel do not evolve significantly, with $G' \sim 5$ kPa and $\tau_{FM} \gtrsim 10^5$s. The elastic modulus dominates over the loss modulus at all measured frequencies; for $\omega = 1 ~\mathrm{\rd{Hz}}$, $G'/G'' \approx 125$.

\subsection{Light scattering}
The small-angle light scattering apparatus is described in detail in~\cite{tamborini_multiangle_2012}. In brief, the scattered light is collected by a lens system and forwarded to the detector of a CMOS camera, such that each pixel corresponds to a well defined scattering vector $\mathbf{q}$, with $q = 4\pi n \lambda^{-1} \sin(\theta/2)$, where $n = 1.338$ is the solvent refractive index, $\lambda = 632.8~\mathrm{nm}$ the laser in-vacuo wavelength and $\theta$ the scattering angle. The $\mathbf{q}$ dependent intensity is obtained as $I(\mathbf{q}) = \left <I_p\right >_\mathbf{q}$, where $I_p$ is the CMOS signal of the $p$-th pixel, corrected for the dark background as in~\cite{duri_time-resolved-correlation_2005}, and $\left <\cdot \cdot \cdot \right >_\mathbf{q}$ is an average over a small region in $q$ space centered around $\mathbf{q}$. For the silica particles used here, the form factor $\approx 1$ in the range of $q$ covered by the setup, such that $I(\mathbf{q})$ is proportional to the static structure factor $S(\mathbf{q})$.

The two-time intensity correlation function is calculated as
$$ g_2(\mathbf{q},t_1,t_2)-1 = \beta \frac{\left < I_p(t_1) I_p(t_2)\right >_\mathbf{q}}{\left < I_p(t_1)\right >_\mathbf{q}\left < I_p(t_2)\right >_\mathbf{q}}-1 \,, $$
where $\beta \gtrsim 1$ is a setup-dependent prefactor chosen such that $g_2(\mathbf{q},t_1,t_2=t_1)-1 = 1$. The intensity correlation function is related to the field correlation function $g_1$ by $g_2-1 = |g_1|^2$~\cite{berne_dynamic_1976}, with $g_1 = F(\mathbf{q},t_1,t_2)/F(\mathbf{q},t_1,t_2=t_1)$ and $F(\mathbf{q},t_1,t_2) = N^{-1}\sum_{j,k=1}^N e^{i\mathbf{q}\cdot[\mathbf{r}_j(t_1)-\mathbf{r}_k(t_2)]}$, where the sum runs over the $N$ particles in the scattering volume.

\subsection{Intermediate scattering function for a purely affine deformation}
Following Ref.~\cite{berne_dynamic_1976} with strain, rather than time, as the independent variable, the intermediate scattering function is expressed as
\begin{equation*}
g_1(\mathbf{q},\Delta \gamma) = \int Q(\Delta \mathbf{r})\exp(-i \mathbf{q} \cdot \Delta \mathbf{r})\mathrm{d} \Delta \mathbf{r} \,,\,\,\,\,\,\,\mathrm{(MM1)}
\end{equation*}
where $Q(\Delta \mathbf{r})$ is the probability distribution function of the particle displacement following a strain increment $\Delta \gamma$. For a purely affine deformation in the direction of $\hat{e}_{\parallel}$, $Q(\Delta \mathbf{r}) = \rd{1/(\Delta \gamma b)}$ and $\Delta \mathbf{r} = \rd{z\Delta \gamma \hat{e}_{\parallel}}$, with $b$ and $z$ the cell gap and the coordinate in the direction of the shear gradient, respectively. By inserting these expressions in Eq. (MM1), one finds $g_1(\qperp, \Delta \gamma) = 1$ and $g_1(\qpar, \Delta \gamma) = \textrm{sinc}\left(q_\parallel \Delta \gamma b/2\right)$. The corresponding $g_2-1$ function is shown as a line in Fig. 2b.

\subsection{Extracting the plastic activity $A(\gamma)$ from the light scattering data}
In order to calculate the plastic activity per unit strain, $A(\gamma)$, we invert Eq. (1) of the main text:
\begin{equation*}
\int_{\gamma_1}^{\gamma_2} A(\gamma) \mathrm{d}\gamma =  \left [ -1\frac{1}{f(\qperp)} \ln \frac{g_1(\qperp,\gamma_1,\gamma_2)}{R(\qperp,\gamma_2-\gamma_1)} \right ]^{\frac{1}{p}} \,.\,\,\,\,\,\,\mathrm{(MM2)}
\end{equation*}
Taking the derivative with respect to $\gamma_2$ at fixed $\gamma_1$ yields
\begin{equation*}
A_{\gamma_1}(\gamma) = f(\qperp)^{-\frac{1}{p}} \left. \frac {\partial}{\partial \gamma_2} \left[ - \ln \frac{g_1(\qperp,\gamma_1,\gamma_2)}{R(\qperp,\gamma_2-\gamma_1)} \right ]^{\frac{1}{p}} \right |_{\gamma_2 = \gamma} \,,\,\,\,\,\,\,\mathrm{(MM3)}
\end{equation*}
where the $\gamma_1$ index in the l.h.s. of Eq. (MM3) indicates that here $A$ is evaluated using data for a specific value of the initial strain $\gamma_1$. Operationally, we calculate  $A_{\gamma_1}(\gamma)$ for several values of $\gamma_1$, using $g_1 = \sqrt{g_2-1}$ and $R(\qperp,\Delta \gamma) = \exp(-q_{\perp}^2 \left < u_{na}^2 \right > /3)$, with $<u^2_{na}> = u^2_{\infty}\frac{\Delta \gamma^2}{\Delta \gamma^2 + \gamma_c^2}$ and $u^2_{\infty} = \rd{0.12} \um^2$, $\gamma_c = \rd{0.12}\%$ [line in Fig. 2d of the main text]. The derivative in the r.h.s. of Eq. (MM3) was performed either numerically on the raw data, or analytically on a 2nd order polynomial fit of $\ln (g_1/R)$. We find similar results and use the latter method, which is less sensitive to data noise. Finally, $A(\gamma)$ is obtained by averaging $A_{\gamma_1}(\gamma)$ over different choices of $\gamma_1$, in the range $1 \% \le \gamma_2 - \gamma_1 \le 6 \%$. The exponent $p$ is chosen by repeating the calculation of $A(\gamma)$ for several test values, finally retaining the $p$ value that minimizes the rms residuals between the experimental
$g_2-1$ and the correlation functions calculated from $A(\gamma)$ using Eq. (1) of the main text.
}

\showmatmethods 

\acknow{This work was funded by ANR (grant n. ANR-14-CE32-0005-01), CNES, and the EU (Marie Sklodowska-Curie ITN Supolen, Grant No.
607937). We thank C. Ligoure, K. Martens, T. Divoux, and D. Vlassopoulos for discussions.}

\showacknow 


\newpage

\newpage
\section{Supplemental information}
\subsection{Setup and sample geometry}
The sample is confined between two glass plates, as shown in Figs. SI-1 and SI-2. The sample volume is 100 $\mu$l, the gap between the plates is 390 $\mu$m. The gel cross section is approximately circular, with a surface $A = 2.6~ \mathrm{cm}^2$, corresponding to an effective radius $\sqrt{A/\pi}=0.91~\mathrm{cm}$. A thin rim of silicon oil is used to seal the lateral surface of the gel, in order to avoid evaporation.
One glass plate is fixed, the other one can slide along the $\hat{\mathbf{e}_{\parallel}}$ direction. The shear stress is applied by means of an electromagnetic actuator; strain is measured by an optical, contactless sensor, as described in [Aime S, et al. (2016) \textit{A stress-controlled shear cell for small-angle light scattering and microscopy}. Review of Scientific Instruments 87(12):123907.]. A laser beam with $1/e^2$ diameter = 2 mm and propagating along the $\hat{\mathbf{e}_{z}}$ direction illuminates the sample. The scattered light is collected by a series of lenses (omitted in Fig. SI-1 for clarity, a full scheme of the scattering apparatus is given in Ref. [2]), such that a CMOS camera records the speckle pattern formed by the scattered light in the far field. The transmitted beam is removed by a beam block. More details on the scattering apparatus are given in [Tamborini E, Cipelletti L (2012) \textit{Multiangle static and dynamic light scattering in the intermediate scattering angle range}. Review of Scientific Instruments 83(9):093106.].
Figure SI-1 shows schematically the setup and the scattering geometry. The intensity correlation functions are obtained from
$$g_2 (\mathbf{q},t_1,t_2 )-1=\beta \frac{\left <I_p (t_1 ) I_p (t_2 )\right>_\mathbf{q}}{\left <I_p (t_1 )\right>_\mathbf{q}\left <I_p (t_2 )\right>_\mathbf{q}}-1$$,
with $\beta \gtrsim 1$ a normalization constant (see Materials and Methods), $I_p(t)$ the scattered intensity at time $t$ for the $p$-th pixel and $\left <\cdot \cdot \cdot \right>_\mathbf{q}$ the average over a set of pixel corresponding to a small region in $q$ space, centered around the scattering vector $\mathbf{q}$. Figure SI-1 shows two such sets of pixels, associated to the same scattering angle $\theta$, and thus the same magnitude of the scattering vector, but with different azimuthal orientation. The region highlighted in green corresponds to the scattering vector $\qpar$ shown in Fig. SI-2, parallel to the shear direction. The region highlighted in blue corresponds to the scattering vector $\qperp$ shown in Fig. SI-2, parallel to $\hat{\mathbf{e}_{\perp}}$ and perpendicular to the shear direction. Intensity correlation functions measured for $\mathbf{q} = \qperp$ are sensitive only to non-affine motion, while correlation functions for $\mathbf{q} = \qpar$ are dominated by affine deformation.

\begin{figure}
\centering
\includegraphics[width=\linewidth]{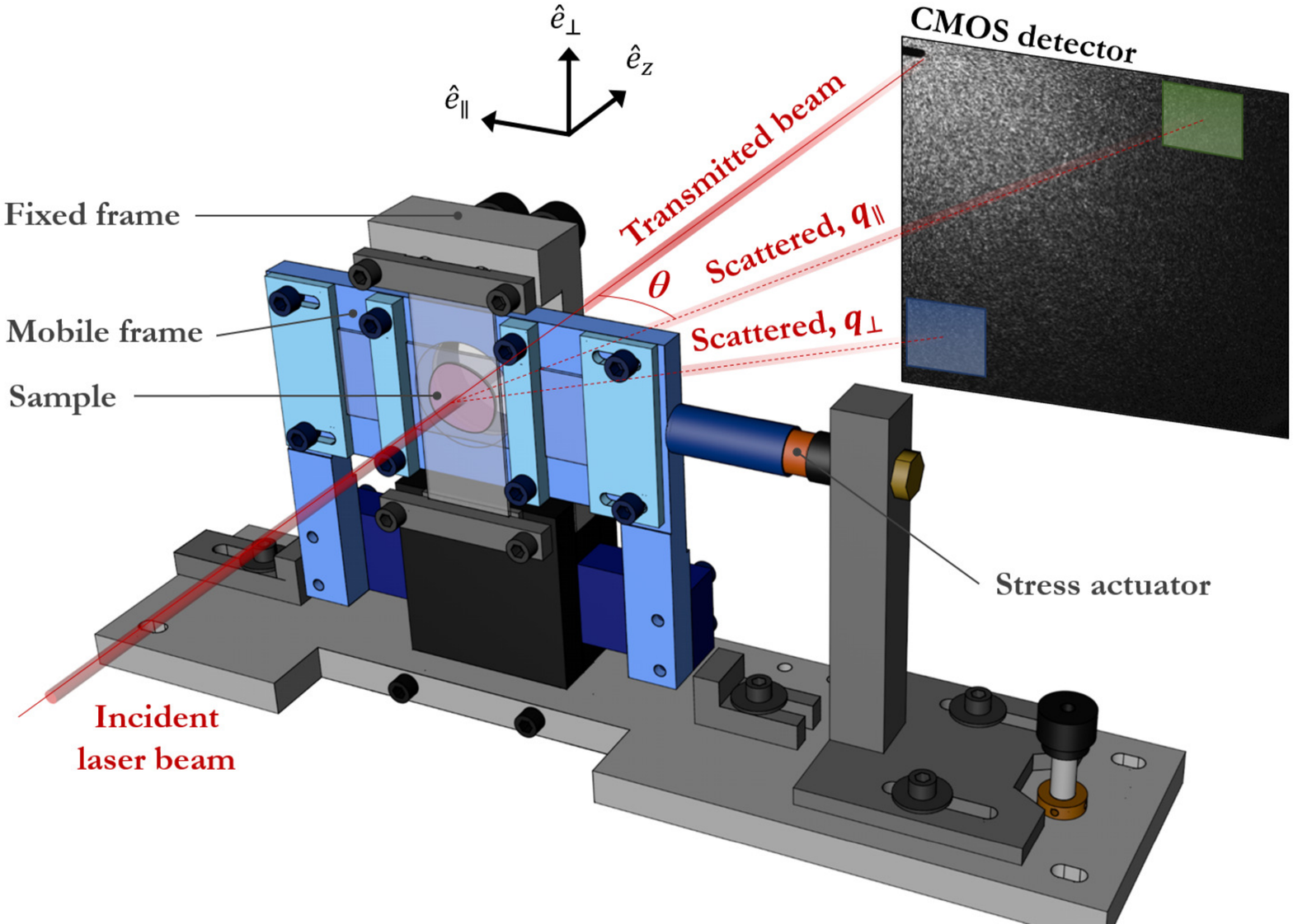}
\caption*{\textbf{Figure SI-1.} Schematic view of the shear cell coupled to the small angle light scattering apparatus. For the sake of clarity, the lenses used to image the far-field scattering intensity on the detector plane are not represented. See text for more details.}
\end{figure}

\begin{figure}
\centering
\includegraphics[width=\linewidth]{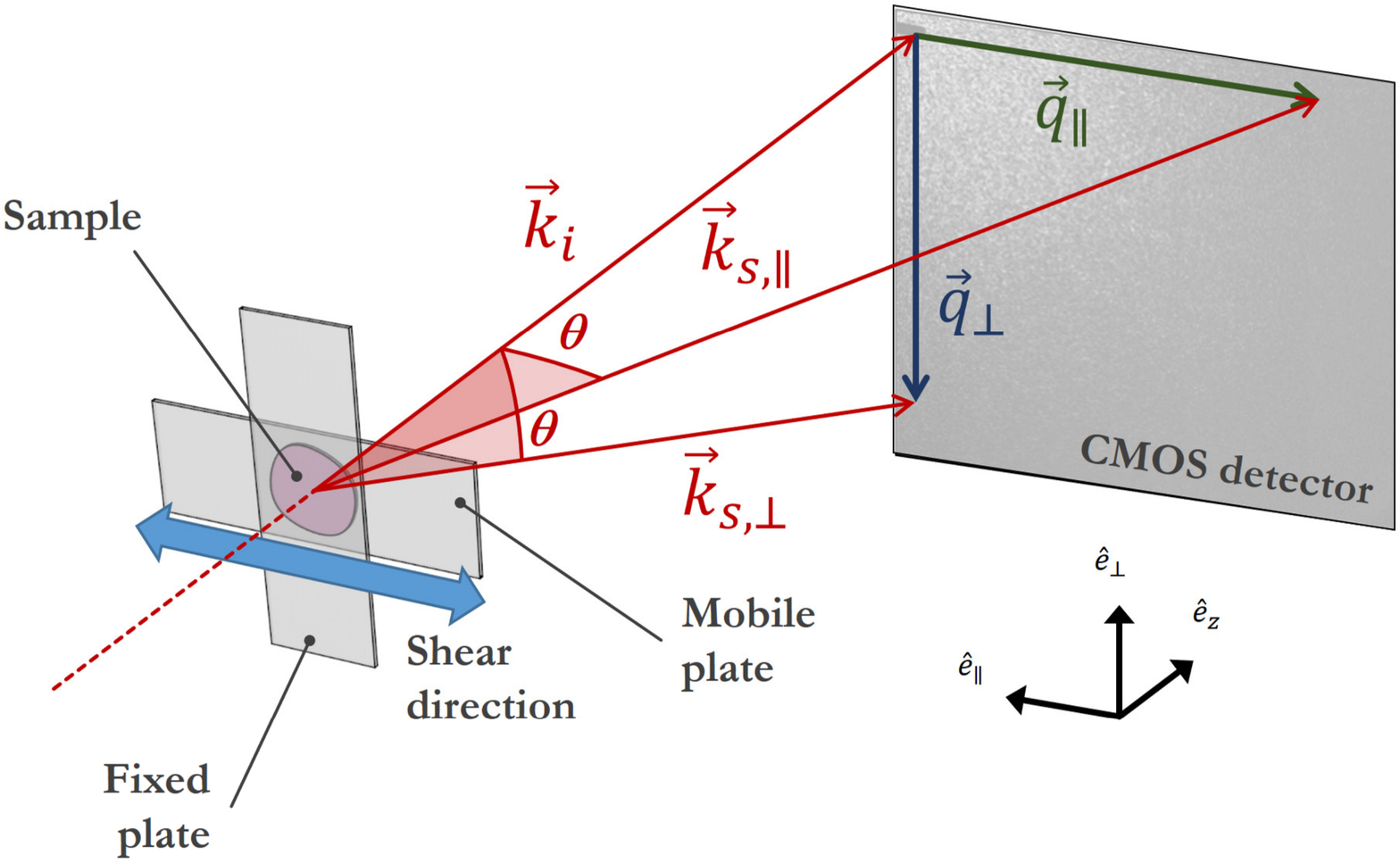}
\caption*{\textbf{Figure SI-2.} Geometry of the light scattering experiment. A shear stress is applied by displacing the mobile plate in the $\hat{\mathbf{e}_{\parallel}}$ direction, as shown by the double arrow. $\mathbf{k}_i$ is the wave vector of the incident light, $\mathbf{k}_{s,\parallel}$ and $\mathbf{k}_{s,\perp}$ are the wave vectors of the light scattered at the same angle $\theta$, but corresponding to scattering vectors oriented along the shear direction and perpendicular to it, respectively. See text for more details.}
\end{figure}

\subsection{Reproducibility}
We have performed eight creep experiments coupling rheology and light scattering and five complementary rheology tests on samples with the same composition as that described in the main text. The rheological properties are very well reproducible, e.g. the elastic modulus of all samples follows the same age dependence to within 25\%. All samples exhibited delayed failure under creep. The failure time varies strongly (from 1h to 50h) for a modest variation of the applied stress, $3.5\% \le \sigma_0⁄G \le 5.5\%$. The general trend is for the failure time to decrease with $\sigma_0$, although large run-to run variations, up to a factor of about 5 are seen for comparable $\sigma_0/G$  values. Both the strong run-to-run variation of the failure time and its marked dependence on $\sigma_0$  have been reported for other soft solids (see e.g. Refs. 24, 27 of the main manuscript). In the five experiments (out of eight) where the failure time exceeded 6h, a dynamic precursor was clearly seen, with the same features as described in the main text: the plastic activity goes through a maximum that corresponds to the minimum of the macroscopic shear rate. One test was performed in a different optical layout, allowing a spatial map of the plastic activity over the full sample to be measured. This test suggests that plasticity is to some extent spatially localized, which may explain why the dynamic precursor was occasionally not seen [In the layout of Fig. SI 1, the laser beam illuminates only about 1\% of the sample.]

\subsection{Comparison between mechanical and dynamical precursors of failure}
Figure 3 of the main manuscript suggests that the onset of plastic activity detected by measuring the microscopic dynamics is approximately concomitant with a deviation of $\dot{\gamma}$ with respect to its initial power law trend. To further investigate the relationship between microscopic and macroscopic quantities, we plot in Fig. SI-3 the strain dependence of the cumulated plastic activity and the relative deviation of $\dot{\gamma}$ from a power law. The former is defined as
$$\frac{\int_0^{\gamma}A(\gamma ')d\dot{\gamma '}}{\int_0^{\gamma_{max}}A(\gamma ')d\dot{\gamma '}}$$ ,
with $\gamma_{max} = 28.4\%$ the strain at which macroscopic failure starts. The relative deviation of $\dot{\gamma}$ is defined as $(\dot{\gamma}-\dot{\gamma}_{FM})/$, with $\dot{\gamma}_{FM}$ the power law fit to the strain rate shown as a dashed line in Fig. 3 of the main text. Both quantities are defined such that they tend to one at large strain. Figure SI-3 shows that both quantities fluctuate around zero (due to noise) at the beginning of the experiment; they start growing significantly when the strain attains about 15\%.
Figure SI-4 shows the cumulated plastic activity plotted against the relative deviation of $\dot{\gamma}$. At the onset of plasticity, the data follow approximately a power law with exponent 3. This demonstrates that the growth of the plastic activity and the deviation of $\dot{\gamma}$ occur indeed concomitantly. Moreover, it demonstrates that the growth of the microscopic plasticity is sharper than that of the deviation of $\dot{\gamma}$, as suggested by the comparison of the two panels of Fig. SI-3. This indicates that $A$ is more sensitive than the strain rate as a signal of plasticity.

\begin{figure}
\centering
\includegraphics[width=\linewidth]{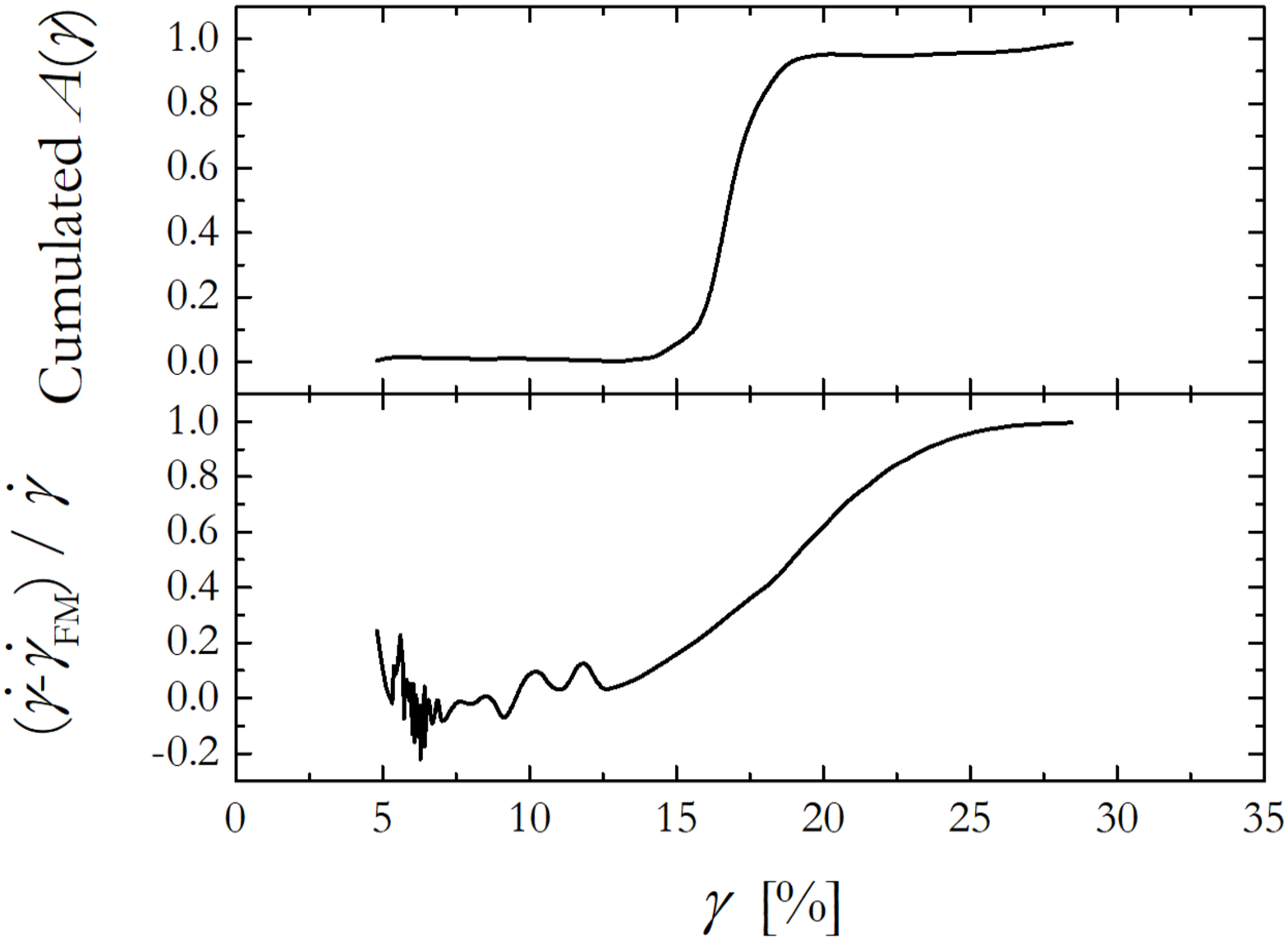}
\caption*{\textbf{Figure SI-3.} Top panel: strain dependence of the cumulated plastic activity. Bottom panel: strain dependence of the deviation of the strain rate with respect to a power law fit to $\dot{\gamma}$ in the first regime of the gel creep (dashed curve in Fig. 3 of the main text).}
\end{figure}

\begin{figure}
\centering
\includegraphics[width=\linewidth]{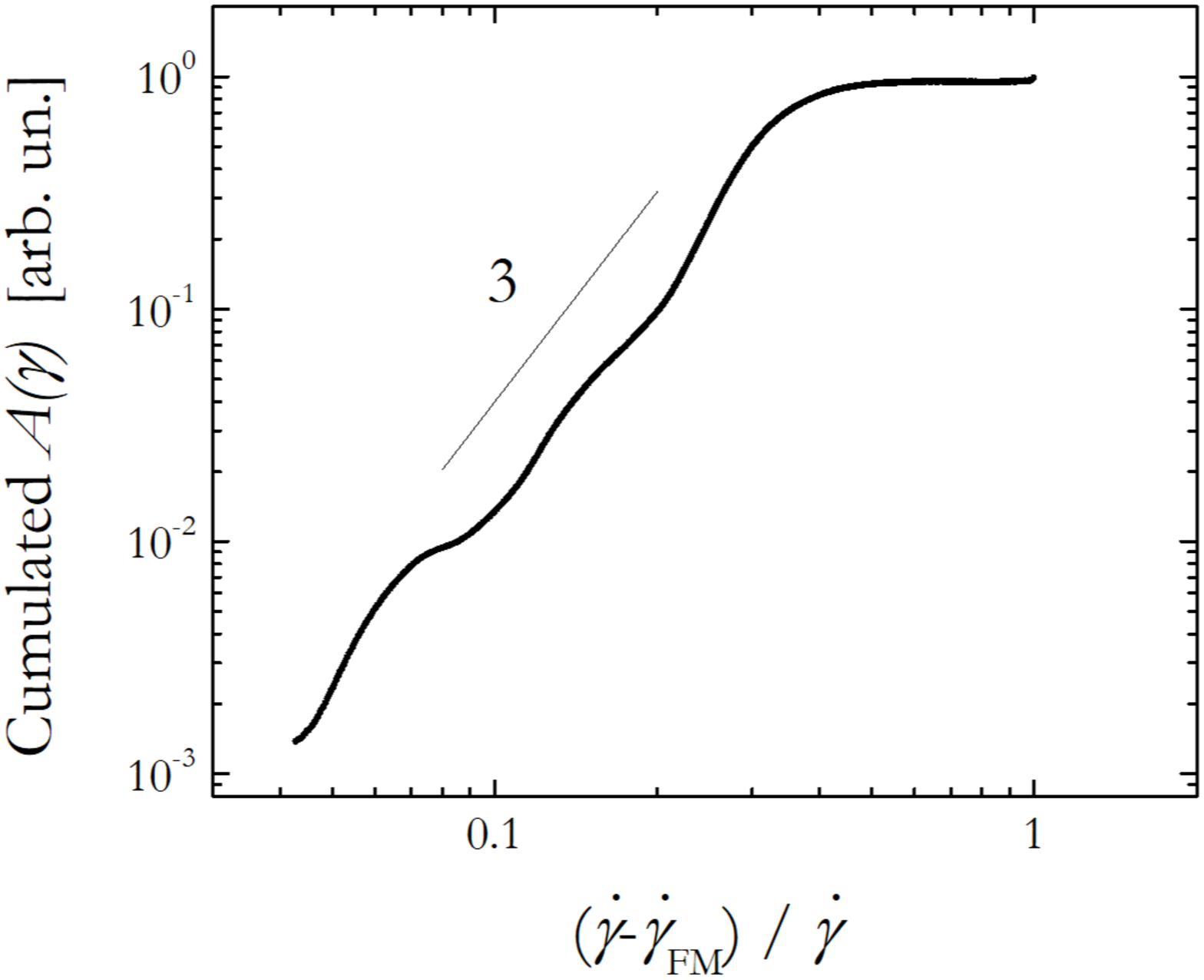}
\caption*{\textbf{Figure SI-4.} Double logarithmic plot of the cumulated plastic activity vs the deviations of the strain rate from a power law. See text for more details.}
\end{figure}

\end{document}